\documentclass[aps, reprint, superscriptaddress, prl, footinbib]{revtex4-1}

\usepackage{amsmath}
\usepackage{amsfonts}
\usepackage{amssymb}
\usepackage{graphicx}
\usepackage[usenames, dvipsnames]{color}
\usepackage{natbib}

\definecolor{mygray}{gray}{0.6}

\newcommand{\ket}[1]{| #1 \rangle}
\newcommand{\bra}[1]{\langle #1 |}

\begin{document}

\title{Nonlinear Polariton Fluids in a Flatband Reveal Discrete Gap Solitons}

\author{V. Goblot}
\email{valentin.goblot@c2n.upsaclay.fr}
\affiliation{Centre de Nanosciences et de Nanotechnologies (C2N), CNRS, Universit\'{e} Paris-Sud, Universit\'{e} Paris-Saclay, 91120 Palaiseau, France}

\author{B. Rauer}
\affiliation{Centre de Nanosciences et de Nanotechnologies (C2N), CNRS, Universit\'{e} Paris-Sud, Universit\'{e} Paris-Saclay, 91120 Palaiseau, France}
\affiliation{Vienna Center for Quantum Science and Technology, Atominstitut, TU Wien, Stadionallee 2, 1020 Vienna, Austria}

\author{F. Vicentini}
\author{A. Le Boit\'{e}}
\affiliation{Universit\'{e} de Paris, Laboratoire Mat\'{e}riaux et Ph\'{e}nom\`{e}nes Quantiques, CNRS, F-75013, Paris, France}

\author{E. Galopin}
\author{A. Lema\^itre}
\author{L. Le Gratiet}
\author{A. Harouri}
\author{I. Sagnes}
\author{S. Ravets}
\affiliation{Centre de Nanosciences et de Nanotechnologies (C2N), CNRS, Universit\'{e} Paris-Sud, Universit\'{e} Paris-Saclay, 91120 Palaiseau, France}

\author{C. Ciuti}
\affiliation{Universit\'{e} de Paris, Laboratoire Mat\'{e}riaux et Ph\'{e}nom\`{e}nes Quantiques, CNRS, F-75013, Paris, France}

\author{A. Amo}
\affiliation{Universit\'e de Lille, CNRS, UMR 8523 -PhLAM- Physique des Lasers Atomes et Mol\'ecules, F-59000 Lille, France}

\author{J. Bloch}
\affiliation{Centre de Nanosciences et de Nanotechnologies (C2N), CNRS, Universit\'{e} Paris-Sud, Universit\'{e} Paris-Saclay, 91120 Palaiseau, France}

\date{\today}

\begin{abstract}
	
	Phase frustration in periodic lattices is responsible for the formation of dispersionless flatbands. The absence of any kinetic energy scale makes flatband physics critically sensitive to perturbations and interactions. We report on the experimental investigation of the nonlinear response of cavity polaritons in the gapped flatband of a one-dimensional Lieb lattice. We observe the formation of gap solitons with quantized size and abrupt edges, signature of the frozen propagation of switching fronts. This type of gap solitons belongs to the class of truncated Bloch waves, and had only been observed in closed systems up to now. Here the driven-dissipative character of the system gives rise to a complex multistability of the flatband nonlinear domains.
	These results open up interesting perspective regarding more complex 2D lattices and the generation of correlated photon phases.
	
\end{abstract}


\maketitle

Geometric frustration in periodic quantum media is responsible for the existence of energy bands with no dispersion. These systems extremely sensitive to any perturbation like disorder or interaction. Fascinating many body physics occurs in flatbands, among which the formation of spin liquids and spin ices \cite{Balents2010}, itinerant ferromagnetism \cite{Lieb1989, Tasaki1992}, fractional quantum Hall states \cite{Tsui1982} or superconductivity in twisted bilayer graphene~\cite{Cao2018, Cao2018b}. In the quest for emulation of this rich phenomenology in controlled systems, pioneering works on frustrated lattices with a flatband have been realized in recent years~\cite{Leykam2018}, using different analog systems like cold atoms \cite{Jo2012, Taie2015}, arrays of coupled optical waveguides \cite{Guzman2014, Mukherjee2015, Vicencio2015} or semiconductor microcavities \cite{Masumoto2012, Jacqmin2014, Baboux2016, Klembt2017, Whittaker2018}. However, most of these works have been limited so far to the linear regime, where particle-particle interactions are negligible. Despite interesting theoretical predictions \cite{Vicencio2013, DiLiberto2018}, the experimental exploration of many-body physics in synthetic frustrated flatbands remains at its infancy.

Exciton-polaritons in semiconductor microcavities have emerged as a powerful platform to study quantum fluids in a driven-dissipative context~\cite{Carusotto2013}. Polaritons are quasi-particles arising from the strong radiative coupling between photons confined in a semiconductor microcavity and excitons confined in quantum wells (QW). Their excitonic component provides significant repulsive interactions (equivalent to a Kerr-like nonlinearity), and the dissipative nature of the system allows for direct injection of polaritons at a given energy. This has enabled the observation of hydrodynamic features including superfluidity~\cite{Amo2008}, nucleation of vortices~\cite{Nardin2011, Sanvitto2011} and solitons~\cite{Grosso2011, Amo2011}. Additionally, the potential landscape probed by polaritons can be sculpted using a variety of techniques~\cite{Schneider2017}. This allows controlling the polariton band structure by engineering lattices and offers a versatile playground to investigate the interplay between kinetic and interaction energy.
Recently several groups have reported the realization of a flatband for polaritons: fragmentation of a polariton condensates induced by disorder was observed~\cite{Baboux2016}, as well as interesting polarization textures induced by spin-orbit coupling~\cite{Whittaker2018}.

In this Letter, we report the experimental investigation of the nonlinear response of polaritons in the flatband of a one-dimensional Lieb lattice of coupled micropillar cavities. We observe the generation of bright quantized nonlinear domains with abrupt and well-defined edges. These sharp profiles are due to the fact that the interaction energy induced by the drive cannot be accommodated as kinetic energy in the flatband. As a result, propagation of switching fronts is frozen. The size of the domains evolves through abrupt jumps as the pumping power is swept, and  multistability is evidenced around each of these jumps. Theoretical analysis of the observed features indicates that these domains belong to the family of gap solitons named truncated Bloch waves~\cite{Alexander2006, Wang2009, Anker2005, Bennet2011, Bersch2012}, which had never been observed in a driven-dissipative context.

\begin{figure}[t]
	\includegraphics[width=\linewidth]{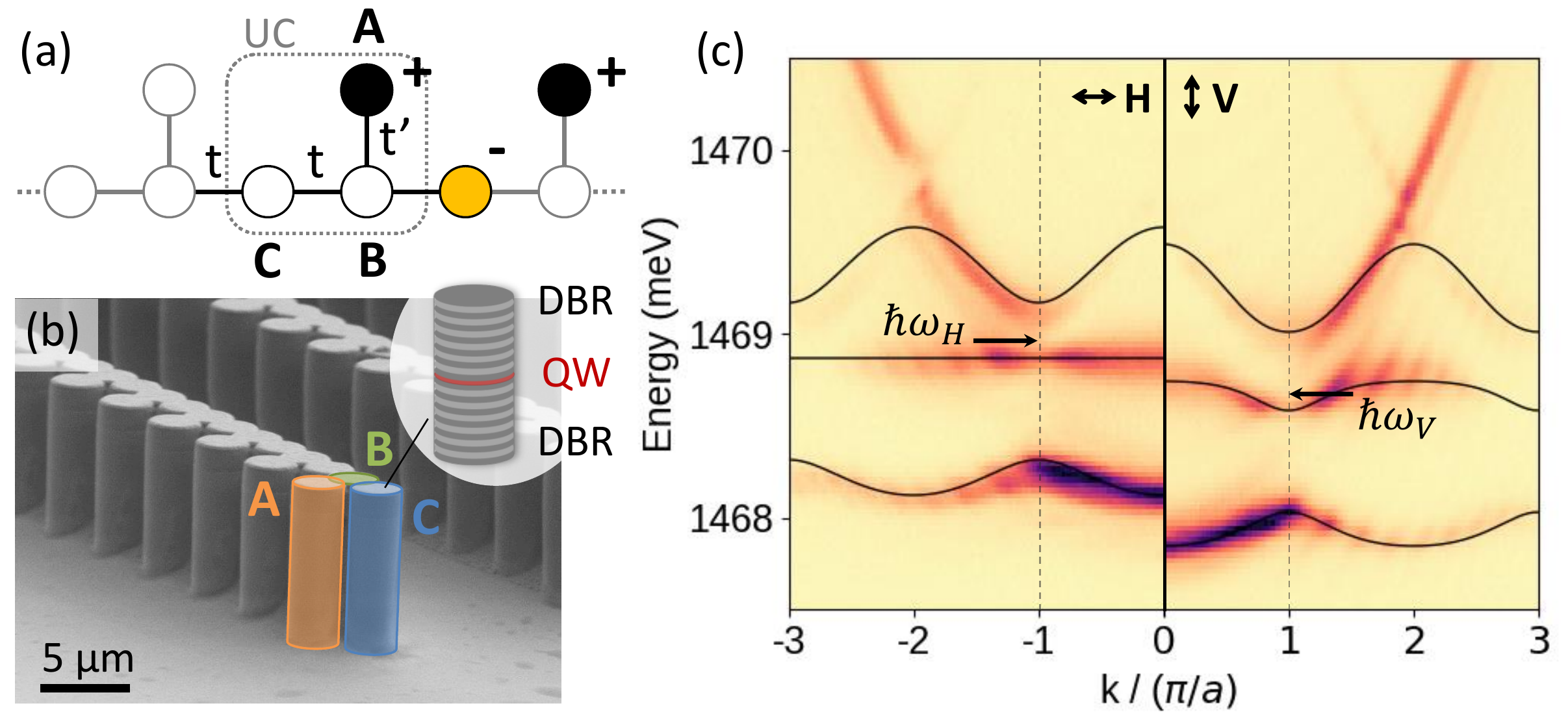}
	\caption{\label{fig1} (a) Schematic representation of the 1D Lieb lattice. Filled circles: localized eigenstate with relative phase indicated with signs. (b) Scanning electron microscopy image of a micropillar lattice. Inset: schematic representation of a single pillar with embedded layers. (c) Energy resolved photoluminescence measured in momentum space under non-resonant pumping, for linear polarization parallel (H) and orthogonal (V) to the lattice.  Solid lines: calculated dispersion solving a tight binding Hamiltonian. For H polarization,  $E_A = E_C = 1468.9 \ \mathrm{meV}$, $E_B = E_C - 0.30\ \mathrm{meV}$ and  $t = t' = 0.30\ \mathrm{meV}$; for V polarization $E_A = E_C + 0.18\ \mathrm{meV}$, $E_B = E_C - 0.30\ \mathrm{meV}$ and $E_C = 1468.6 \ \mathrm{meV}$, with same $t, t'$.
		Arrows indicate the energy $\hbar \omega_{H,V}$ and wave vector of the resonant pump.
	}
\end{figure}

The 1D Lieb lattice (see Fig.~\ref{fig1}(a)) is one of the simplest lattices hosting a flat energy band~\cite{Lieb1989}. The unit cell (UC) contains three -$A, B, C$- sites linked by nearest-neighbor couplings $t$ ($t'$) between $B$ and $C$ ($A$ and $B$) sites.
In the tight-binding approximation, the corresponding single-particle Hamiltonian is:
\begin{align}
\hat{H} = \sum_{l, n} E_l \ket{l_n} \bra{l_n} - \sum_n \Big( t \big( \ket{B_n} \bra{C_n}  \nonumber \\
+ \ket{B_n} \bra{C_{n+1}} \big) + t' \ket{A_n} \bra{B_n} + \mathrm{h. c.}\Big)
\label{eq:Lieb}
\end{align}
where $\ket{l_n}$, with $l\in\{A, B, C\}$, is the state on site $l$ in the $n$th unit cell, with on-site energy $E_l$.
The energy spectrum of the 1D Lieb lattice presents three bands. While eigenfunctions in the lower (resp. upper) band are constructed with the same (opposite) phase for neighboring sites, eigenfunctions for the middle band show alternating phase sign on the $A, C$ sublattice. This creates a phase frustration on sites $B$: when $E_A = E_C$, a destructive interference induces zero wave function amplitude on $B$ sites. As a result, the middle band is flat with localized eigenstates of the form $\ket{f_n} = \left(\ket{A_{n-1}} - \ket{C_n} + \ket{A_n} \right) / \sqrt{3}$, as shown in Fig.~\ref{fig1}(a). When $E_A \neq E_C$, the interference on $B$ sites is not fully destructive and the middle band becomes dispersive.

To experimentally implement a polariton Lieb lattice, we use a semiconductor heterostructure grown by molecular beam epitaxy. It consists of a $\lambda$ GaAs layer embedded between two $\mathrm{Ga_{0.9}Al_{0.1}As/Ga_{0.05}Al_{0.95}As}$ distributed Bragg reflectors (DBR) with 36 (top) and 40 (bottom) pairs (quality factor $Q \approx 50000$). A single 8~nm $\mathrm{In_{0.05}Ga_{0.95}As}$ QW is inserted at the center of the cavity, resulting in a $3.5\ \mathrm{meV}$ Rabi splitting. The cavity is processed into an array of coupled micropillars (Fig.~\ref{fig1}(b)) using electron beam lithography and dry etching down to the GaAs substrate. Each micropillar is mapped to a site of the tight-binding Hamiltonian while the coupling between sites is provided by the finite overlap between neighboring pillars. The on-site energy of the lowest energy mode in each micropillar is controlled by the pillar diameter, but also depends on the number of adjacent pillars, whose proximity reduces the local confinement. Thus a fine tuning in the pillar diameters is required to match $A$ and $C$ on-site energies and obtain a flatband. We choose a 2.4 $\mathrm{\mu m}$ distance between adjacent pillars (resulting in UC size $a = 4.8\ \mathrm{\mu m}$), and $3.0\,\mathrm{\mu m}$, $2.8\,\mathrm{\mu m}$ and $2.9\,\mathrm{\mu m}$ for the diameters of $A$, $B$ and $C$ pillars.

Optical spectroscopy is performed at 4~K, using  a tunable continuous wave (cw) monomode excitation laser focused on the 1D lattice. The signal is collected in transmission geometry through the back of the sample, using a lens with 0.5 numerical aperture, and focused on the entrance slit of a spectrometer coupled to a CCD camera. The emission is analyzed in real- or momentum-space, by imaging either the sample surface or the Fourier plane of the collection lens. Using a $\lambda$/2 wave-plate and a polarizer, we select the linearly polarized emission either parallel (H), or orthogonal (V) to the lattice.

First, we characterize the band structure of the polariton lattice in the linear regime. The lattice is excited non-resonantly tuning the laser energy around 1.6 eV, with a Gaussian-shaped elongated spot and weak pumping power. The energy difference between the uncoupled cavity mode and the exciton resonance amounts to $-3.5\ \mathrm{meV}$. The momentum-space resolved emission for both H and V polarization is shown in Fig.~\ref{fig1}(c). Three bands are evidenced arising from  hybridization of micropillar confined modes. They are well reproduced by tight-binding calculations, except for the upper band where the expected band folding is not observed in the experiment. This deviation can be explained in terms of mixing with higher-energy bands, neglected by the tight-binding description. For H polarization, the middle band is dispersionless and is gapped from the two other bands. Because of polarization dependent boundary conditions for the electromagnetic field, the flatband condition cannot be achieved simultaneously for two orthogonal linear polarizations~\cite{Baboux2016}. Indeed, under V polarization all bands are dispersive. The faint intensity modulation visible in Fig.~\ref{fig1}(c) arises from multiple reflections between the bottom mirror and the polished back side of the substrate.

\begin{figure}[t]
	\includegraphics[width=\linewidth]{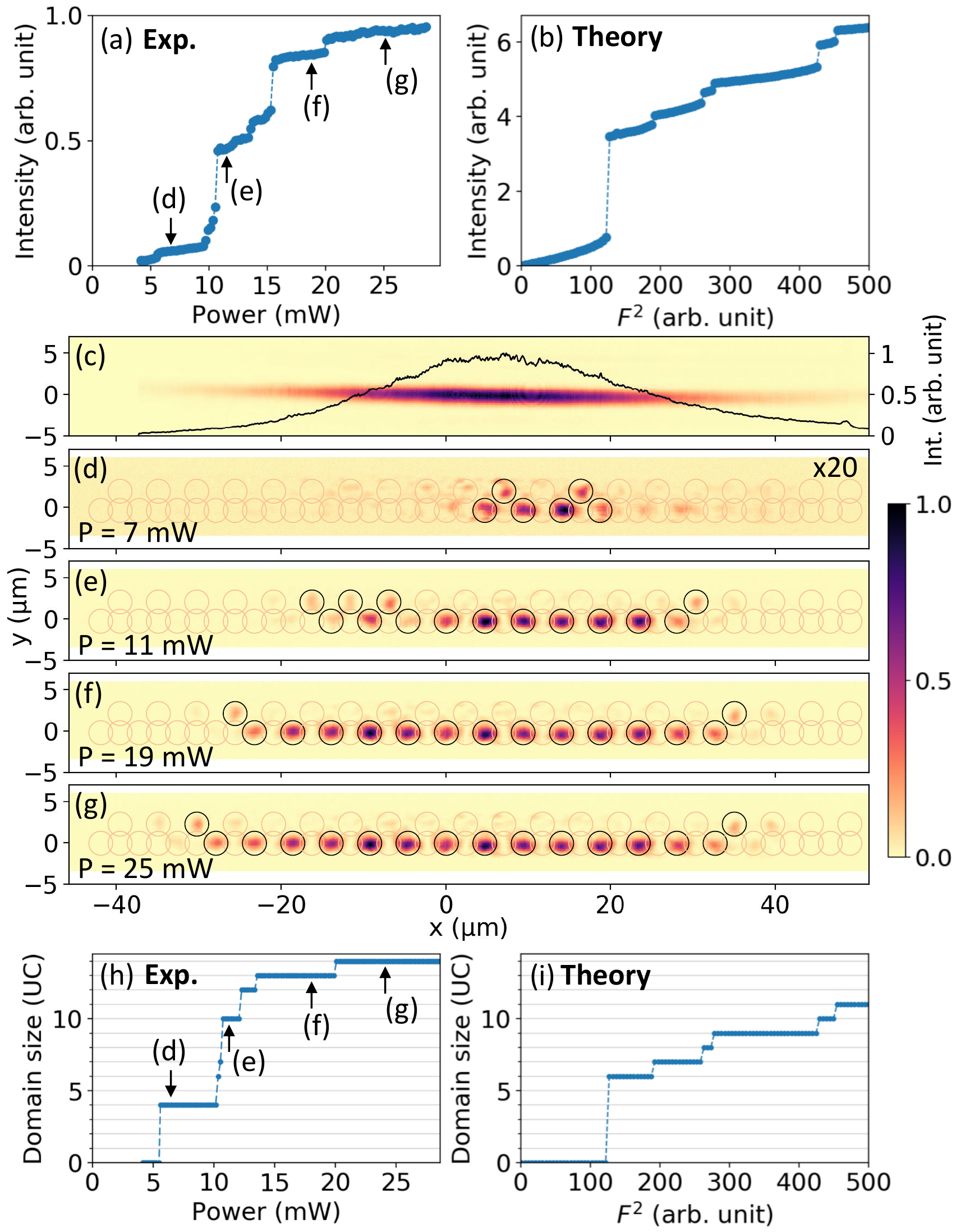}
	\caption{\label{fig2} (a-b): Total emission intensity (a) measured, (b) calculated under resonant excitation tuned to energy $\hbar \omega_H$ as a function of power; (c) 2D real-space image of the excitation spot. Black line: spatial profile integrated over the transverse direction; (d-g) Real-space emission intensity measured at pumping powers indicated in (a). Highly excited micropillars are indicated by circles. (h-i) Size of the nonlinear domains (h) measured and (i) calculated as a function of (h) excitation power, (i) $F^2$.}
\end{figure}

To investigate the polariton nonlinear response in the flatband, we inject polaritons with a quasi-resonant laser at energy detuning $\Delta$ from the flatband. In the experiments described in the following, the flatband is at energy $E_C = 1468.9\ \mathrm{meV}$ and the laser energy $\hbar \omega_H$, is blueshifted by $\Delta = 90~\mathrm{\mu eV}$, corresponding to about one third of the energy gap separating the flatband from the upper one (Fig.~\ref{fig1}(c)). The pumping beam is polarized along H and focused into a Gaussian-shaped elongated spot, of $40\,\mathrm{\mu m} \times 3\,\mathrm{\mu m}$ FWHM (Fig.~\ref{fig2}(c)). Its $5.0^\circ$ angle of incidence matches the edge of the first Brillouin zone (BZ), i.e. $k = \pi/a$, enabling efficient coupling to the flatband modes~\cite{Supplementary}.

Figure~\ref{fig2}(a) presents the measured total emitted intensity when increasing the excitation power $P$. Several abrupt jumps are observed separated by plateaus where the intensity weakly varies. To discuss the origin of the observed features, we show in Fig.~\ref{fig2}(d-g) spatial emission patterns measured for different excitation powers. For $P = 7~\mathrm{mW}$, above the lowest power intensity jump, Fig.~\ref{fig2}(d) evidences the formation of a 4 UCs nonlinear domain. It is located around the center of the excitation spot and its shape does not evolve when $P$ is further increased, up to a power of 10 mW, where the next jump happens. We then observe the formation of a larger nonlinear domain with 10 UCs. Actually, every jump corresponds to an evolution of the domain size by a discrete number of UCs, as summarized in Fig.~\ref{fig2}(h). The size of the domains is well defined because their edges are extremely sharp: the emission intensity drops by more than an order of magnitude over one UC on each side.

To get more insight into the physics, we solve the steady-state of a discretized Gross-Pitaevskii equation that includes pump and loss terms \cite{Carusotto2004}. The evolution of the polariton amplitude $\psi_n$ on site $n$, under a cw resonant drive, is governed, in the frame rotating at the drive frequency $\omega$, by the equation:
\begin{align}
i \hbar \frac{\mathrm{d} \psi_n(t)}{\mathrm{d} t} =& \left(E_n - \hbar \omega + \hbar g |\psi_n (t)|^2 -i \frac{\gamma}{2} \right) \psi_n (t) \nonumber \\
& - \sum_m t_{nm} \psi_m (t) + i F e^{-x_n^2 / 4\sigma^2 } e^{-i k_p x_n}
\label{eq:GPE}
\end{align}
where $\hbar g$ is the polariton-polariton interaction constant, $\gamma$ the polariton linewidth. $E_n$ is the on-site energy and $t_{nm}$ are the couplings to neighboring sites, deduced from $t$ and $t'$ defined before~\cite{Supplementary}. $F$ is the drive amplitude, $k_p = \pi / a$ the mean wavevector of the drive and $x_n$ the spatial position of site $n$. The spatial distribution of the drive excitation is chosen Gaussian with $\sigma = 3.5$ UCs and its detuning from the flatband, defined as $\Delta = \hbar \omega - E_C$, is $\Delta = 3 \gamma = 90\ \mathrm{\mu eV}$.

Figure~\ref{fig2}(b) presents the calculated total intensity in a 40 UC chain when increasing the drive amplitude. A series of abrupt jumps is observed in good qualitative agreement with the experiment. Similar to the experiment, the simulated steady-state spatial profiles consist in nonlinear domains of finite size~\cite{Supplementary} and each jump in the total intensity corresponds to a discrete change in the domain size.

The origin of these quantized nonlinear domains in the flatband is intimately linked to the fact that the pump energy lies in an energy gap. Locally, the system switches into the nonlinear regime (with high polariton intensity) when the local pump intensity is higher than a threshold value so that the local interaction energy $\hbar g |\psi|^2 \gtrsim \Delta$. Outside this high excitation region, the system remains in the linear regime with very low intensity because the excitation laser lies within an energy gap. This is the reason why abrupt edges are formed. The size of the domain is determined by the number of UCs for which the local pump intensity is higher than the threshold value. Because of the Gaussian shape of the excitation spot, the threshold condition is reached for more and more UCs as the power is increased, and domains of increasing size are formed. If the excitation profile was completely flat, all excited UCs would switch simultaneously.

These nonlinear states belong to the general family of gap solitons. They have been discussed in a different context~\cite{Alexander2006, Wang2009} and named truncated Bloch waves (TBW), because their pattern is similar to a spatial portion of the excited Bloch states. Notice that in our system, TBW can also be observed when exciting in a gap above a dispersive band, as illustrated by simulations presented in the supplemental~\cite{Supplementary}. In a flatband, there is no kinetic energy to overcome so that discrete TBW are formed as soon as the laser detuning overcomes the band linewidth.

A significant difference between numerical simulations and experiments is revealed in the quantitative domain size versus $F$. Once the first domain is formed (at $F^2 =120$), the calculated domain size increases by exactly one UC at each jump (see Fig.~\ref{fig2}(i)). The situation is different in the experiment where for instance an increase of 6 UCs is observed at 10 mW (Fig.~\ref{fig2}(h)). In the supplemental~\cite{Supplementary}, we show that disorder in the on-site energies explains this apparent discrepancy. Experimentally, disorder mainly stems from small fluctuations in the pillar size and shape, caused by etching. A local redshift of the flatband eigenstates acts as a barrier for the nonlinear domain: when the excitation strength is sufficient to overcome this barrier, the domain size can abruptly increase by several unit cells, as observed experimentally.
Notice also that the drive with wave vector at the edge of the BZ imposes a $\pi$ phase difference between adjacent UCs. This creates destructive interference on $A$ sites explaining why $A$ sites are generally dark within the nonlinear domains. Bright $A$ sites are only observed at the edges of the nonlinear domains or in regions where disorder is of the order of the interaction energy, thus altering the interference (see for instance Fig.~\ref{fig2}(d) or Fig.~\ref{fig2}(e), left side).

\begin{figure}[t]
	\includegraphics[width=\linewidth] {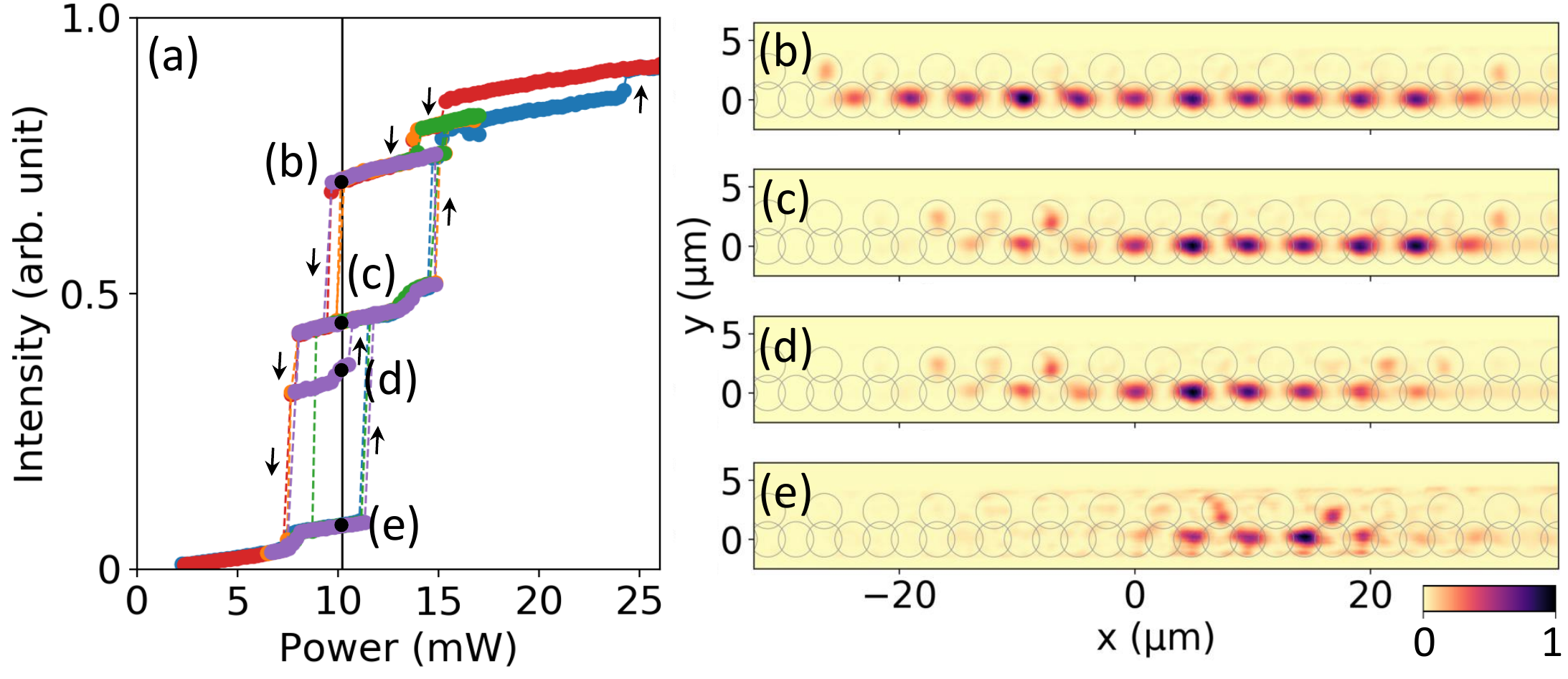}
	\caption{\label{fig3} Total intensity measured under resonant excitation of the flatband (same excitation parameters as in Fig.~\ref{fig2}) when scanning the excitation power up and down as indicated by arrows; (b-e) real space emission patterns measured for $P = 10.2$ mW on different branches as indicated in (a). }
\end{figure}

TBW have been experimentally observed in closed systems~\cite{Anker2005, Bennet2011, Bersch2012}, but to our knowledge have never been reported in a driven-dissipative system. Associated to the presence of dissipation, such nonlinear photonic systems are expected to present hysteretic behaviors~\cite{Carusotto2013, Baas2004}. To probe this property, we excite the flatband with $\Delta = 90\ \mathrm{\mu eV}$, and scan the power up and down around each nonlinear jump of the intensity. This reveals multiple hysteresis cycles, summarized in Fig.~\ref{fig3}(a). In fact, each abrupt jump is associated with a hysteresis cycle. Depending on the value of the power and on the power ramp history, several configurations of nonlinear domains can be achieved. As an illustration, Fig.~\ref{fig3}(b-e) show four nonlinear spatial patterns which can be generated for $P = 10.2$ mW. Each of them shows a TBW soliton, with well defined number of bright unit cells and no emission outside the nonlinear domains. Thus the polariton nonlinear response of the flatband reveals complex multistable behavior when scanning the power up and down.

\begin{figure}[t]
	\includegraphics[width=\linewidth]{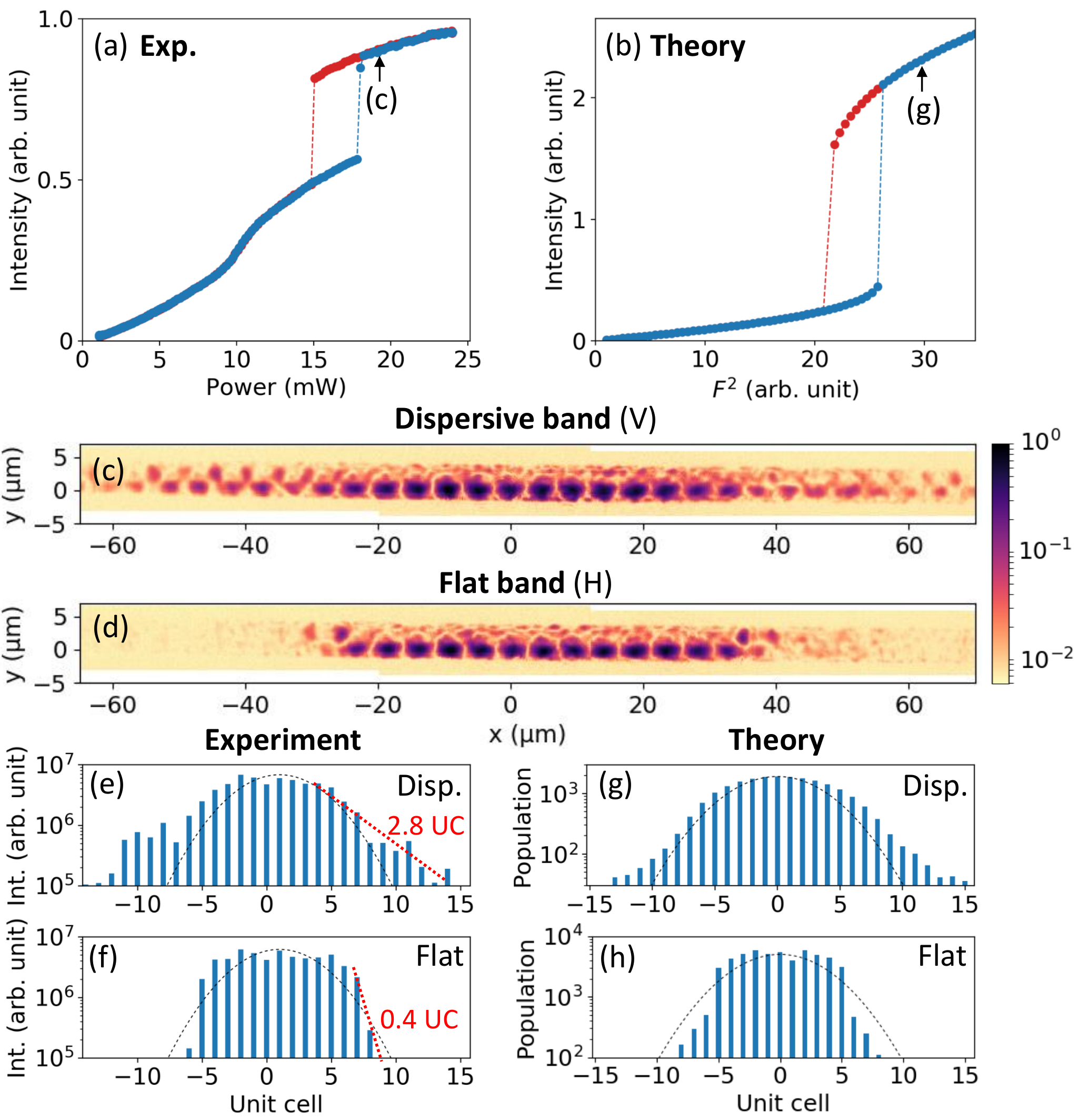}
	\caption{\label{fig4} (a): Total intensity  measured  as a function of excitation power under V polarized excitation tuned to an energy $\hbar \omega_V$. (b) Corresponding calculated intensity as a function of $F^2$. In (a) and (b), blue (resp. red) color corresponds to increasing (decreasing) excitation power. (c) and (d): Spatially resolved emission represented in logarithmic color scale measured for $P = 19~\mathrm{mW}$ (c) in the dispersive (V polarized pump), (d) in the flatband (H polarized pump).  (e-f) Measured integrated intensity on $C$ sites extracted from (c),(d). Black lines: pump spot profile, red line: exponential fit of the emission intensity spatial decay. (g-h): Calculated intensity on C sites for (g) a dispersive ($E_A=6\gamma$) and (h) a flat  ($E_A=0$) band considering (g) $\Delta=3\gamma$, $F^2=300$ and (h) $\Delta=2\gamma$, $F^2=30$.}
\end{figure}

We now compare the experiments described above with the nonlinear response when pumping the system within a dispersive band. Direct comparison can be done in the very same lattice simply by injecting polaritons with an excitation linearly polarized along the V direction. The polarization dependent on-site energies result in a middle band that is dispersive in this case (Fig.~\ref{fig1}(c), right panel). We use pumping conditions comparable to those used in Fig.~\ref{fig2}: $\Delta = 60~\mathrm{\mu eV}$ with respect to the bottom of the band, and same angle of incidence (Fig.~\ref{fig1}(c)). The results are presented in Fig.~\ref{fig4}(a): a nonlinear increase in the total emitted intensity is observed at $P = 10~\mathrm{mW}$, followed by an abrupt intensity jump at $P = 17~\mathrm{mW}$, associated with a hysteresis cycle. The emission spatial pattern measured in the nonlinear regime for $P = 19~\mathrm{mW}$ is shown in logarithmic scale in Fig.~\ref{fig4}(c). Polariton propagation out of the pumping region is clearly evidenced: the emission spreads over the entire portion of the lattice under investigation, in stark contrast with the case of the flatband (also shown in logarithmic scale in Fig.~\ref{fig4}(d)). Exponential fits to the intensity profiles (see Fig.~\ref{fig4}(e,f)) allow estimating a polariton propagation distance of $13.6~\mathrm{\mu m}$ (2.8 UCs) for the dispersive band, to be compared to only $2.1~\mathrm{\mu m}$ (0.4 UCs) for the flatband. Polariton propagation outside the high density region is possible in the dispersive band thanks to the conversion of interaction energy into kinetic energy. The resulting propagation of switching fronts has been previously used to realize a spin switch~\cite{Amo2010}. When injecting the quantum fluid into a gap, such propagation is totally suppressed.

We simulated the resonant excitation experiment for the case of the dispersive band. The calculated total intensity shown in Fig.~\ref{fig4}(b) presents a single hysteresis cycle, in agreement with previous reports~\cite{Carusotto2013, Baas2004}. In the supplemental, we show that the first nonlinear increase observed in the experiment at $P = 10~\mathrm{mW}$ is well accounted for by introducing disorder in the simulation~\cite{Supplementary}. The spatial profiles calculated for the dispersive and the flatband with parameters corresponding to Fig.~\ref{fig4}(e,f) (no disorder) are shown in Fig.~\ref{fig4}(g,h). They nicely reproduce the difference in intensity profile when propagation of the switching front is frozen (flatband) or not (dispersive band).

In conclusion, we have shown that the nonlinear response of a polariton fluid resonantly injected into a flatband is governed by the emergence of gap solitons of the family of TBW. They are discrete nonlinear domains whose abrupt edges reflect the freezing of kinetic energy, and show complex multistable patterns under non-homogeneous spatial excitations.
Similar experiments in 2D lattices with geometric frustration, where the flatband is touching a dispersive band (kagome) or crosses Dirac points (2D Lieb lattice), could offer interesting perspectives, as well as the investigation of the Bogoliubov excitations when the flatband is driven in the nonlinear regime.
Using polariton structures with stronger interaction strength~\cite{Togan2018, Rosenberg2018, Tsintzos2018, Knueppel2019}, experiments beyond semi-classical approximation could be envisioned: in this context, flatbands are also particularly relevant for the study of many-body correlated phases~\cite{Wu2007, Sun2011, Neupert2011, Clark2019}.

\begin{acknowledgments}
	The authors are grateful to F. Baboux, P. St-Jean and G. Malpuech for fruitful discussions.
	This work was supported by the H2020-FETFLAG project PhoQus (project no. 820392), the QUANTERA project Interpol (ANR-QUAN-0003-05), the French National Research Agency (ANR) project Quantum Fluids of Light (ANR-16-CE30-0021), the Labex NanoSaclay (ICQOQS, Grant No. ANR-10-LABX-0035), the French RENATECH network, the ERC via the Consolidator Grant CORPHO No. 616233 and the Austrian Science Fund (FWF) through the doctoral program CoQuS (W1210) (B.R.), the Labex CEMPI (ANR-11-LABX-0007), the CPER Photonics for Society P4S, the I-Site ULNE  (projet NONTOP) and the M\'etropole Europ\'eenne de Lille (project TFlight)”
\end{acknowledgments}


%

\clearpage

\onecolumngrid

\setcounter{figure}{0} 
\renewcommand{\thefigure}{S\arabic{figure}}

\subsection{\large Supplemental Material: Nonlinear Polariton Fluids in a Flatband Reveal Discrete Gap Solitons}

\section{Resonant excitation of the flatband modes}
\begin{figure}[!h]
	\includegraphics[width=0.6\linewidth]{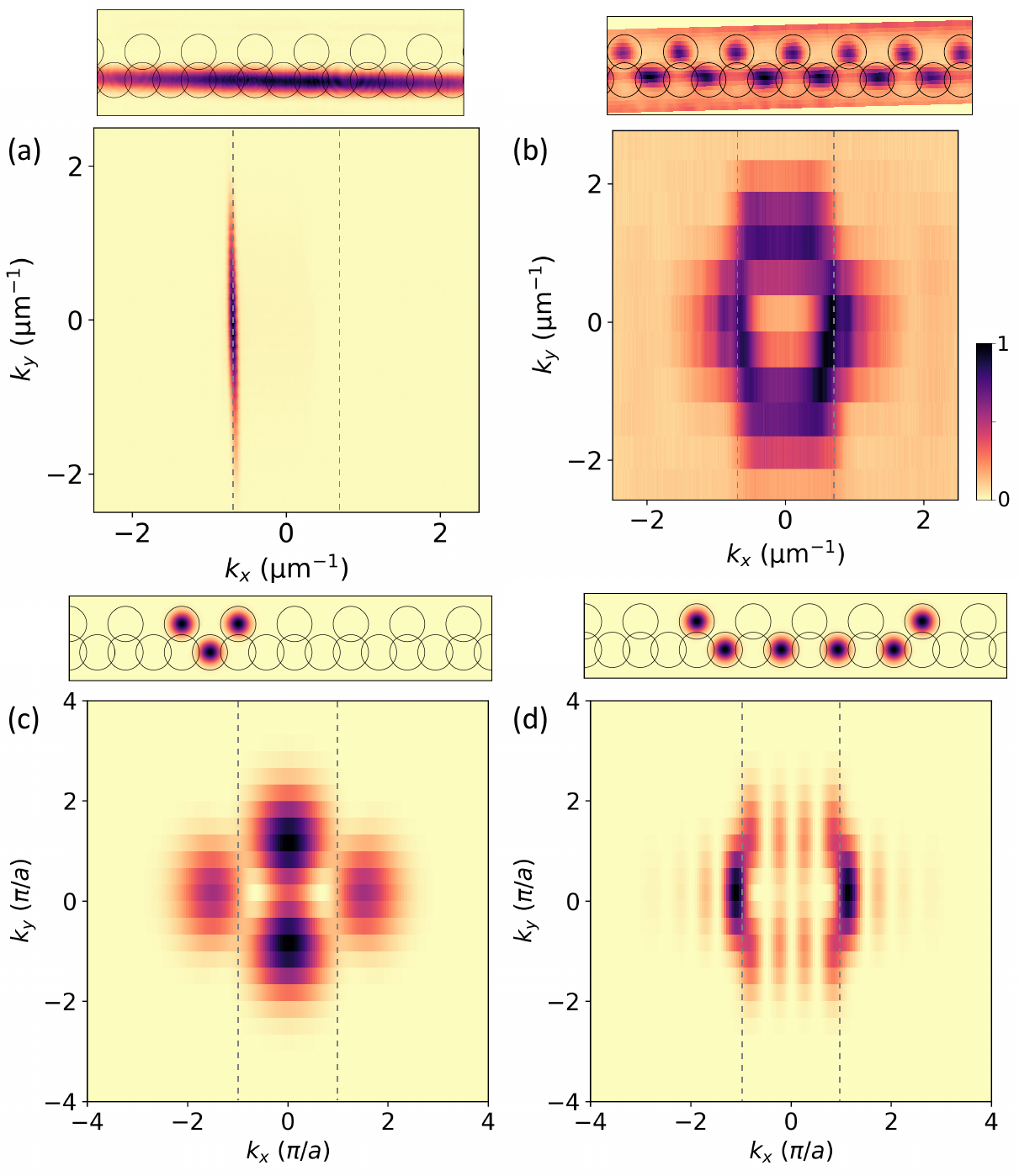}
	\caption{\label{figS1} (a) Real- (top) and momentum-space (bottom) images of the laser spot used for resonant excitation. (b) Measured real- and momentum-space photoluminescence at the flatband energy. (c,d) Calculated real- and momentum-space emission pattern of two examples of flatband eigenstates: (c) a single plaquette $\ket{f_n}$; and (d) a linear combination of four plaquettes with alternating sign on neighboring cells $\sum_{j=0}^{3}(-1)^j \ket{f_{n+j}}$. In all panels, dotted lines indicate the edges of the first Brillouin zone.
	}
\end{figure}

To determine the optimal excitation scheme for the resonant drive of the flatband, we study the momentum-space resolved photoluminescence (PL) of the flatband modes. This pattern can be obtained by spectrally filtering the emission, measured under weak non-resonant pumping, at the flatband energy. The $(k_x, k_y)$ map of the emission is then reconstructed from spectra measured at different values of $k_y$, such as the one shown in Fig. 1(c) of the main text. The result is presented in Fig.~\ref{figS1}(b): the intensity is zero at the center of the Brillouin zone (BZ) ($(k_x, k_y) = 0$), and is maximal at the BZ edges ($k_x = \pi/a$). This reflects the antisymmetric nature of the flatband eigenmodes (opposite phase on $A,C$ pillars). Thus, to ensure efficient coupling to the polariton states in the flatband, the pumping beam is tilted from normal incidence by $5.0^\circ$ along the $x$ direction, corresponding to the edge of the first BZ.

The real-space PL at the energy of the flatband can also be obtained with a similar method (spectral filtering of the real-space PL) and is shown at the top of Fig.~\ref{figS1}(b). As expected due to geometric frustration, we measure vanishing intensity on $B$ sites.

Let us comment briefly on the consequence of exciting the flatband at the BZ edge. When the wave-vector of the driving field is equal to $k=\pi/a$, it imposes a phase difference of exactly $\pi$ between neighboring unit cells. The opposite sign on $C$ sites in neighboring unit cells leads to a destructive interference on $A$ sites. As an illustration, Fig.~\ref{figS1}(c,d) shows calculated  real- and momentum-space emission pattern of two different localized eigenstates: a single plaquette $\ket{f_n}$, and a linear superposition of four plaquettes, of same magnitude but alternating sign on neighboring unit cells, which can be written $\sum_{j=0}^{3}(-1)^j \ket{f_{n+j}}$. To compute these radiation patterns, we use a simplistic description of the eigenfunctions of the chain of pillars: we consider a Gaussian-shaped orbital per pillar (corresponding to the $s$ mode).
To construct a given wave function, we assign the amplitude and phase computed from the tight-binding model to each of these Gaussian shaped orbitals.
The momentum-space radiation pattern is obtained by Fourier transformation of this wave function. The state in Fig.~\ref{figS1}(d), with a phase difference of $\pi$ between neighboring plaquettes, corresponds to a Bloch state $\ket{\psi(\pi/a)}$, but truncated to only 4 unit cells.

Importantly the non-linear domains measured in our experiment indeed present a spatial pattern similar to the one calculated in Fig.~\ref{figS1}(d). This pattern reflects the phase imposed by the drive at the edge of the BZ, resulting in low intensity on $A$ sites inside the non-linear domains because of destructive interferences. $A$ sites have significant intensity only at the edge of the domains or in regions where disorder overcomes interaction energy (like in Fig.~2(d) or Fig.~2(e) in the main text) and locally breaks the destructive interferences.

\section{Parameters for numerical simulations}

The discrete Gross-Pitaevskii equation introduced to model our experiments (Eq. (2) of the main text) is a set of $3N$ equations, that describe the time evolution of the polariton amplitude on each site. $N$ is the number of unit cells in the lattice. The coupling terms $t_{n,m}$ between the different sites are linked to the coupling constant $t$ and $t'$ of the Lieb tight-binding Hamiltonian (Eq. (1) of the main text) as follows:

\begin{align}
t_{nm} =
\begin{cases}
t &\mathrm{if\ sites\ } n,m \mathrm{\ are\ neighboring\ } B \mathrm{\ and\ } C \mathrm{\ sites,}\\
t' &\mathrm{if\ sites\ } n,m \mathrm{\ are\ neighboring\ } A \mathrm{\ and\ } B \mathrm{\ sites,}\\
0 &\mathrm{otherwise,\ i.e.\ if\ sites\ } n,m \mathrm{\ are\ not\ neighbors.}
\end{cases}
\end{align}

The values of the parameters are deduced from the tight-binding fit to the measured polariton dispersion as shown in Fig. 1(c) of the main text. In the case of the flatband, we take $E_A = E_C = E_0$, $E_B = E_C - 10 \gamma$ and $t = t' = 10 \gamma$, with $\gamma = 30 \mathrm{\mu eV}$ (and the energy offset $E_0 = 0$ for simplicity). For the dispersive band, we use $E_A = E_C + 6\gamma$, all other parameters being unchanged.

\section{Numerical simulations: steady-state spatial profiles}

\begin{figure}[!h]
	\includegraphics[width=0.7\linewidth]{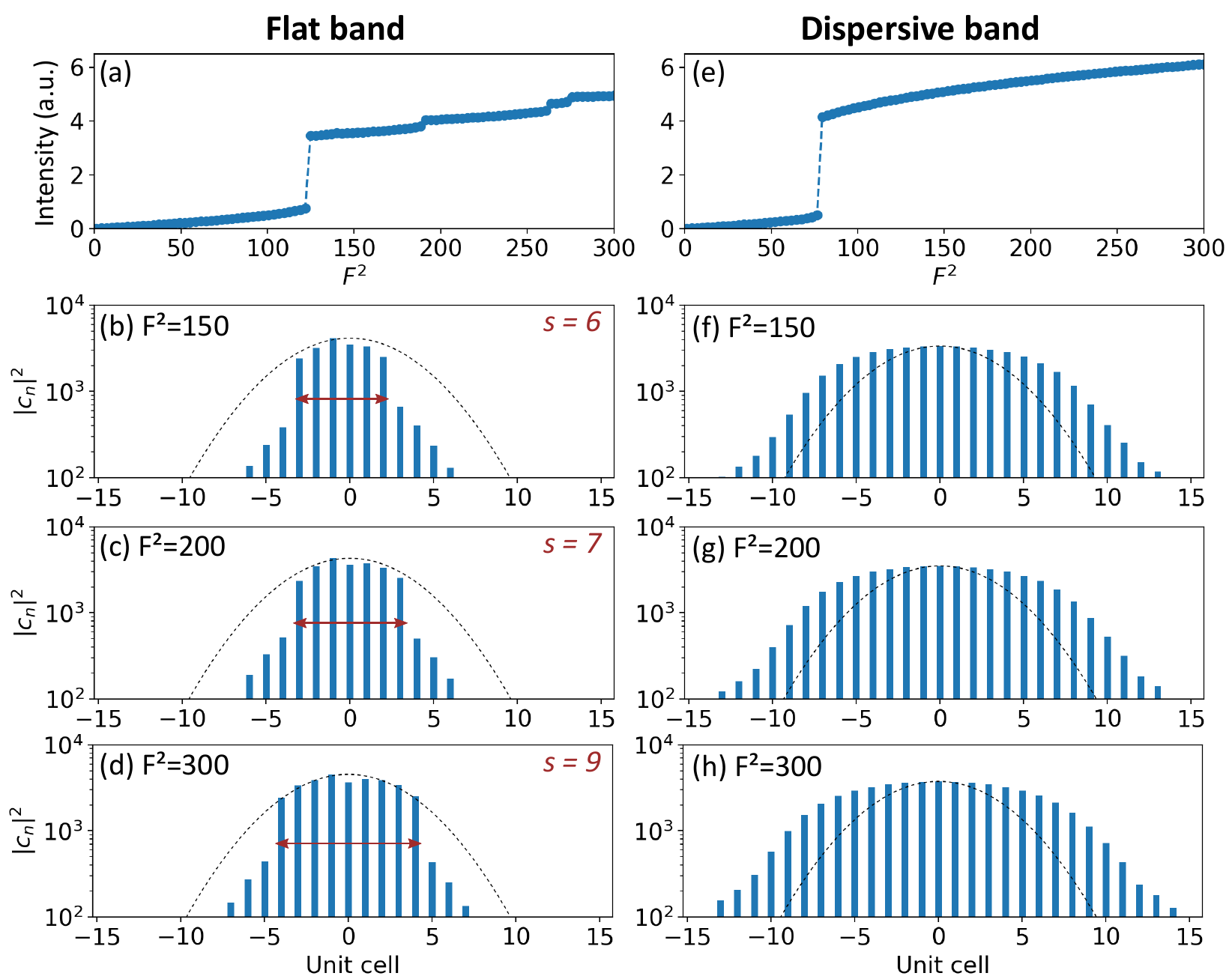}
	\caption{\label{figS2} \textbf{Comparison between the polariton non-linear dynamics in the flat and when driving the system within a dispersive band} :  (a-d) calculation for the flatband with $E_A = 0$; (e-h) calculations for the dispersive band with $E_A = 6 \gamma$; in both cases $\Delta = 3 \gamma$. (a,e) Total intensity calculated as a function of $F^2$. (b-d, f-h) Steady-state occupation $|c_n|^2$ on sites $C$ for various values of the drive intensities $F^2$. Dashed lines indicate the shape of the pumping spot. For panels (b-d), $s$ is the domain size.
	}
\end{figure}

In this section we present the steady-state spatial intensity profiles calculated for the quasi-resonant injection of polaritons in the flatband, that correspond to the total intensity and domain size from Fig.~2(b,i) of the main text. The drive detuning is $\Delta = 3 \gamma$ and the drive wave vector $k_p = \pi/a$. The calculated total intensity versus drive power $F^2$ from Fig.~2(b) of the main text is reproduced in Fig.~\ref{figS2}(a). Fig.~\ref{figS2}(b-d) shows the calculated spatial profile on pillars $C$ for different values of $F^2$. In each case, a nonlinear domain delimited by a sharp drop in occupation at the edges is clearly identified. As $F^2$ is increased above the first abrupt intensity jump, each intensity jump corresponds to an increase of the domain size by exactly 1 unit cell (UC).

As a comparison, we repeat the numerical simulation with same excitation conditions ($\Delta = 3 \gamma$, $k_p = \pi/a$), but with $E_A = 6 \gamma$, such that the middle band is now dispersive and the drive frequency lies within the band. The results are presented in Fig.~\ref{figS2}(e-h). As detailed in the main text, a single jump is observed in the calculated total intensity as the drive power $F^2$ is increased. Moreover, the spatial density profile in the nonlinear regime is smooth and does not evolve significantly as $F^2$ is increased.

\clearpage

\section{Influence of disorder in a flatband}

\begin{figure}[!h]
	\includegraphics[width=0.5\linewidth]{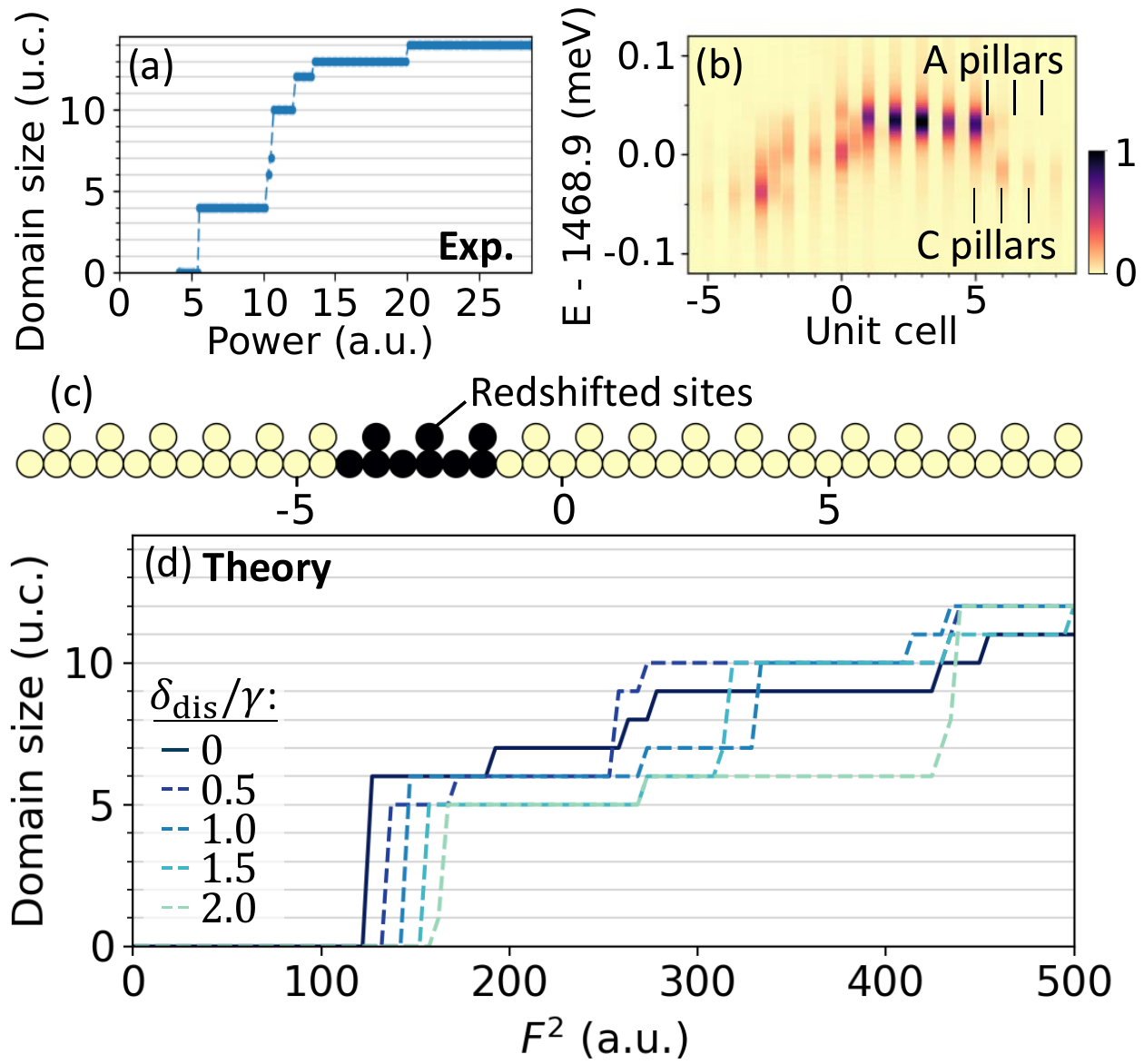}
	\caption{\label{figS3} \textbf{Influence of disorder on the nonlinear regime for the flatband.} (a) Measured size of the nonlinear domains as a function of excitation power (reproduced from Fig. 2(h) of the main text). (b) Measured intensity profile on pillars $A,C$ as a function of resonant drive energy in the linear regime ($P=10 \mathrm{\mu W}$). (d) Calculated size of the nonlinear domain for increasing excitation drive $F^2$, when including a redshift $\delta_{dis}$ on the sites indicated in black in (c). This redshift mimics the effect of local disorder in the chain.
	}
\end{figure}

As explained in the main text, we experimentally observe that the size of the nonlinear domain formed in the flatband can jump by several unit cells at a time when the pumping power is increased (see Fig. 2(h) of the main text, reproduced in Fig.~\ref{figS3}(a)). This feature is not reproduced by the simulation (see Fig. 2(i) of the main text), where increases by one unit cell only are obtained (except for the first jump).
In the following we show that disorder in the on-site energies can explain this discrepancy.

Disorder can strongly affect the physics of particles in a flatband, for example leading to the fragmentation of a bosonic condensate into plaquette-sized localized modes \cite{Baboux2016}. Indeed, since kinetic energy is zero in a flatband, any finite amount of disorder will break the flatband picture. In a dissipative context, disorder strength needs to be greater than the linewidth to significantly alter the physics. Experimentally, disorder mainly stems from small fluctuations in the pillar size and shape, caused by etching.

An estimate of the disorder strength can be extracted from resonant spectroscopy of the flatband eigenstates in the linear regime. Figure~\ref{figS3}(b) shows the light intensity transmitted through $A$ and $C$ pillars when scanning the laser energy. When the laser is in resonance with an eigenstate, an intensity maximum is observed. The figure clearly indicates some spatial energy spreading of the eigenstates across the lattice. More precisely for this particular part of the chain, a redshift is observed to the left of UC 0, and to the right of UC 5. The maximal energy difference between the different states is around $80\, \mathrm{\mu eV}$, comparable to the laser detuning $\Delta = 90 \mathrm{\mu eV}$ used in Fig.~\ref{figS3}(a) (and Fig. 2 of the main text). Thus in the experiments disorder strength is comparable to the interaction energy.
Note that imaging of the eigenstates as done in Fig.~\ref{figS3}(b) does not allow extracting precise on-site disorder on each individual pillar.
Nevertheless, to get a flavor of the effect of disorder on the nonlinear domains, we  introduce in our simulations a distribution of on-site energies which results in a distribution of eigenstates resembling the measured one. A redshift $\delta_{\mathrm{dis}}$ is introduced for the on-site energy of all sites on 2 UCs to the left of the excitation spot (see Fig.~\ref{figS3}(c)). The corresponding simulation of the nonlinear domain size versus $F^2$ is shown in Fig.~\ref{figS3}(d) for different disorder strengths. As in the experiment, we observe series of jumps of different amplitudes.
For instance for $\delta_{\mathrm{dis}} = 0.5 \gamma$, an abrupt jump from 6 to 9 UCs occurs at $F^2 \approx 250$. It corresponds to a progression of the domain edge through all redshifted sites at once. For stronger disorder amplitude, additional big jumps in the domain size are observed at higher excitation powers. The redshifted sites thus create a barrier for the domain edges, and modify the growth of the domains with power.
In the experiment, the disorder landscape is certainly more complex but this simple simulation provides good understanding of the effect of disorder on the observed nonlinear dynamics. We have verified on several lattices realizing different disorder configurations that the nonlinear behavior reported here is qualitatively the same.

\section{Truncated Bloch Waves in the gap above a dispersive band}

\begin{figure}[!h]
	\includegraphics[width=0.7\linewidth]{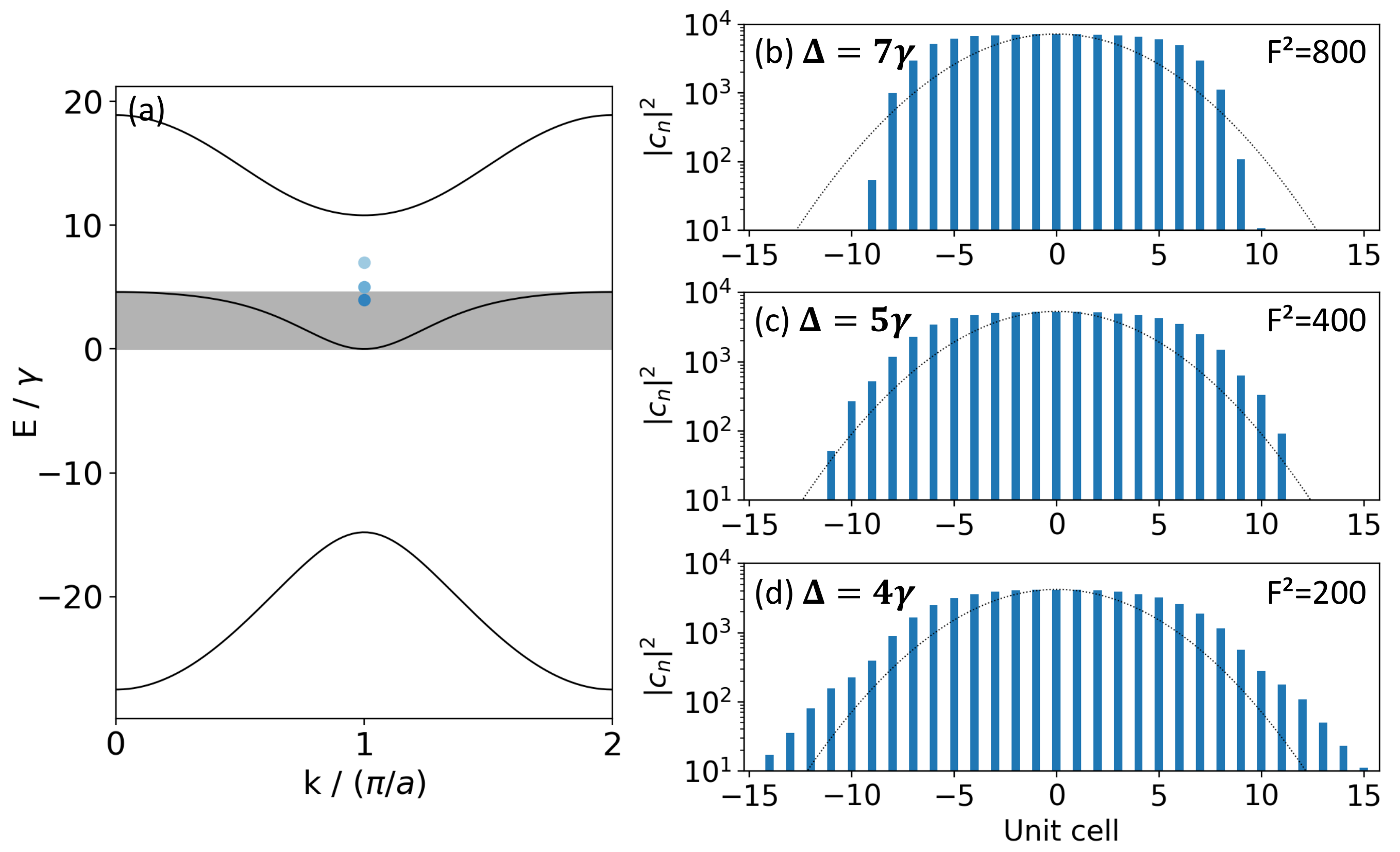}
	\caption{\label{figS5} (a) Band structure calculated by diagonalization of the tight-binding Lieb Hamiltonian for $E_A = 6 \gamma$, $E_B = -10 \gamma$ and $E_C = 0$. The shaded gray region indicates the total spectral width of the middle band. Blue dots indicate the drive energy and wave vector used in panels (b-d). (b-d) Steady-state occupation $|c_n|^2$ on sites $C$ calculated for different values of $\Delta$, in the nonlinear regime (for a value of $F^2$ indicated in each panel).
	}
\end{figure}

We investigate with numerical simulations the behavior of a nonlinear fluid injected in the gap above a dispersive band. Figure~\ref{figS5}(b-d) presents the steady-state profiles calculated in the nonlinear regime and without disorder, for $E_A = 6 \gamma$ and different values of the drive energy detuning with respect to the bottom of the middle band: $\Delta = 4$, $5$ and $7 \gamma$. Note that for $E_A=6 \gamma$, the width of the middle band is $\sim 4.6 \gamma$, so that when $\Delta > 4.6 \gamma$ the drive lies within the gap. For $\Delta = 4 \gamma$, i.e. for a drive below the band edge, the propagation outside the spot is visible as a spatial exponential decay of the intensity. The propagation length $L$ characteristic of the spatial decay is given by $L = v_g / \gamma$, with $v_g = \hbar^{-1} (\partial E / \partial k)$ the group velocity at energy $\Delta$. Increasing the drive energy to $\Delta = 5 \gamma$, a sharp spatial decrease in the intensity is now observed at UC $\pm 11$. For $\Delta = 7 \gamma$, further into the gap, the domain edge is even sharper. In this excitation configuration, since the drive injects polaritons within the gap, there is no single-particle state at this energy. As a result, the interaction energy provided by the drive cannot be converted into kinetic energy: propagation of particles out of the excitation region is prevented. This localization mechanism arising from the interplay between interactions and the existence of an energy gap is precisely the one at play in the formation of gap solitons, and in particular of Truncated Bloch Waves, as originally discussed in Refs.~\cite{Alexander2006, Alexander2006b, Wang2009}.

Thus when the dispersive band is excited within the gap at high energy, Truncated Bloch waves are excited in a similar way as for the flatband. The energy injected in the system is larger than the maximum kinetic energy the system can accommodate so that non linear domains with sharp edges are formed. In the flatband, since kinetic energy is strictly zero, this regime is achieved as soon as the driving energy overcomes the other energy scales of the system, namely the spectral linewidth and disorder.

\section{Multistability of the nonlinear domains}

\begin{figure}[!h]
	\includegraphics[width=0.6\linewidth]{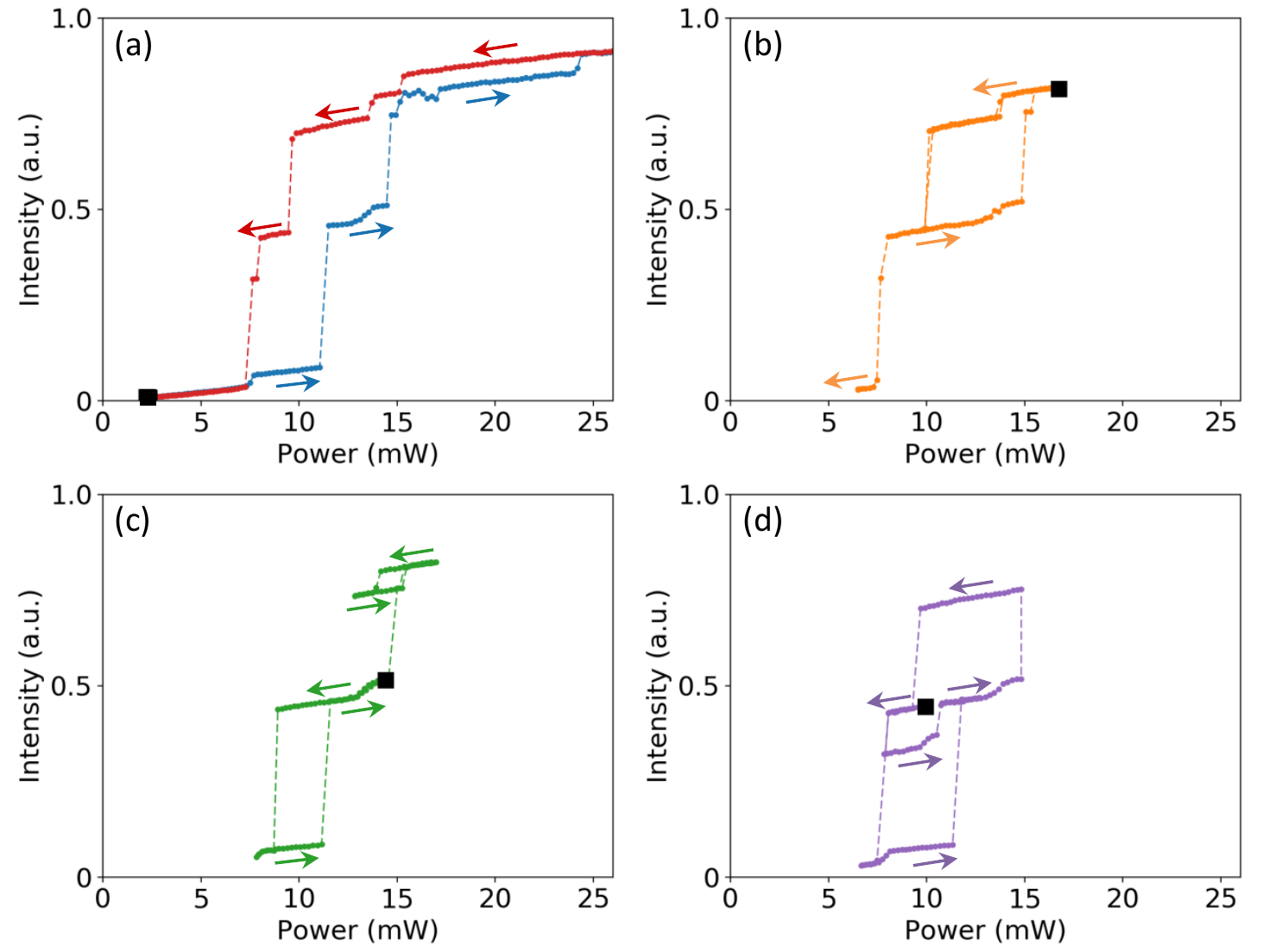}
	\caption{\label{figSxx} (a-d) Total emission intensity measured under resonant excitation of the flatband for different the power scans. In each panel, the starting excitation condition is denoted by a black square and arrows indicate the scan direction.
	}
\end{figure}

In Fig.~\ref{figSxx} we present several experimental power scans obtained with excitation parameters similar as those used in Fig.~2  of the main text ($\Delta = 90\ \mathrm{\mu eV}$ and $k_p = \pi / a$). For each of these power scans, the starting condition is denoted by a black square. Fig.~3(a) of the main text reproduces all these measurements on top of each other.

\section{Influence of disorder within a dispersive band}

\begin{figure}[!h]
	\includegraphics[width=0.7\linewidth]{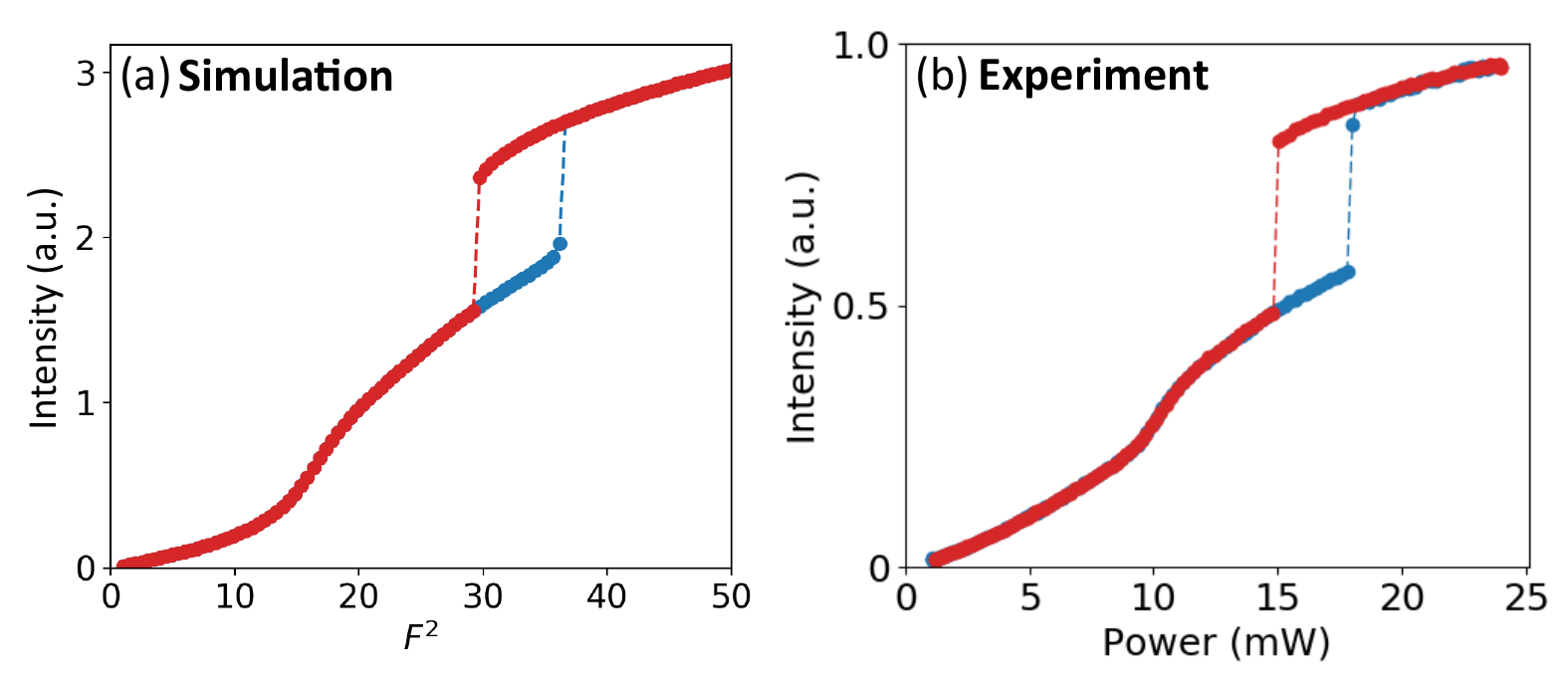}
	\caption{\label{figS4} \textbf{Influence of disorder on the nonlinear regime for the dispersive band.} (a) Calculated total intensity in the lattice versus $F^2$, for an increasing (blue) and decreasing (red) drive intensity. The redshift amplitude is $\delta_{\mathrm{dis}} = 2 \,\gamma$ on the same sites as for the flatband, and the drive detuning $\Delta = 1.5 \gamma$. (b) Total emission intensity measured in the dispersive band as a function of excitation power (reproduced from Fig.~4(a) from the main text).
	}
\end{figure}

Disorder also has an influence on the nonlinear regime in the dispersive band. This is due to the fact that the disorder amplitude in the experiment, on the order of $80\ \mathrm{\mu eV}$, is comparable to the interaction and kinetic energy of the fluid with our choice of laser detuning $\Delta = 60\ \mathrm{\mu eV}$.
In Fig.~\ref{figS4}, we present the results of a numerical simulation taking into account disorder in the dispersive band: we introduce a redshift $\delta_{\mathrm{dis}} = 2 \gamma$ on the same sites as in Fig.~\ref{figS3}(c), $\Delta= 1.5 \gamma$ and $E_A = 6 \gamma$. The total population versus $F^2$ in the up and down scans are in excellent agreement with the experimental results from Fig.~4(a) of the main text, reproduced in Fig.~\ref{figS4}(b). Indeed, the presence of disorder explains the first nonlinear increase in the total intensity before the abrupt jump (only one jump was observed in disorder free simulations, see Fig.~4(b) of the main text).

\end{document}